# On the spectral unfolding of chaotic and mixed systems

Sherif M. Abuelenin*

*Department of Electrical Engineering, Faculty of Engineering, Port-Said University, Port-Fouad, Port-Said, 42526, Egypt. s.abuelenin@eng.psu.edu.eg

## Abstract

Random matrix theory (RMT) provides a framework to study the spectral fluctuations in physical systems. RMT is capable of making predictions for the fluctuations only after the removal of the secular properties of the spectrum. Spectral unfolding procedure is used to separate the local level fluctuations from overall energy dependence of the level separation. The unfolding procedure is not unique. Several studies showed that statistics of long-term correlation in the spectrum are very sensitive to the choice of the unfolding function in polynomial unfolding. This can give misleading results regarding the chaoticity of quantum systems. In this letter, we consider the spectra of ordered eigenvalues of large random matrices. We show that the main cause behind the reported sensitivity to the unfolding polynomial degree is the inclusion of specific extreme eigenvalue(s) in the unfolding process.

## 1. Introduction

Random matrix theory (RMT) provides a framework to studying the relation between quantum systems and their classical counterparts which have chaotic dynamics [1, 2]. RMT models a chaotic system by an ensemble of random Hamiltonian matrices that depends only on the symmetry properties of the system. RMT was successfully applied in the study of the structure of networks arising from real world systems [3-7]. One of the fundamental signatures of quantum chaos is the conjectured link between the statistical fluctuations of the energy spectrum and the integrability or chaotic properties of the Hamiltonian [8-12].

RMT aims to understand correlations between energy levels independently of the variation of level spacing. To study the statistical fluctuations of the energy spectrum it is common to "unfold" the spectrum by means of a transformation [13] involving the cumulated level density, so that the mean level spacing is equal to one. The removal of the influence of the level density is often done by calculating the cumulative spectral function as the number of levels below or at the level $E$. This is frequently referred to as the staircase function. It can be separated into an average (smooth) part $N_{avg}(E)$, whose derivative is the level density, and a fluctuating part $N_{fluc}(E)$. $N_{avg}(E)$ is calculated for the matrix by running spectral average. If the functional form of the mean level density $\rho(E)$ is known, we obtain;

$$I(E) = \int_{-\infty}^{E} dE' \rho(E').$$

In most of practical applications, the exact form of the cumulated level density is not known. The averaged staircase function has to be smoothed or predicted. It follows that unfolding is not a unique procedure. This introduces ambiguities in the unfolding procedure [14]. The unfolding details are usually omitted in the literature [9], but a common method of unfolding is done by arbitrarily parametrizing (or fitting) the numerically obtained level density in terms of a polynomial of degree $n$ [8], often with $n = 3$. After extraction of the average part $I_{avg}(E)$, it is unfolded from the spectra by the introduction of a dimensionless variable

$$\epsilon_i = I_{avg}(E_i)$$

In this variable, the spectra have unity mean level spacing everywhere. Thereafter, the observables calculated for the unfolded spectrum for each matrix are averaged over the ensemble. This is in the same spirit as it was done in spectra of nuclei [15]. These spectra were unfolded for each nucleus separately.

Because of the ambiguity of the unfolding, some standard unfolding procedures spoil the spectral statistics [16]. While under-fitting can enhance fluctuations, over-fitting can remove a great portion of them. This can lead to making erroneous conclusions about the chaoticity of the studied system [9].

It was reported that the statistics that measure long-range correlations, namely the spectral rigidity and the level number variance, strongly depend on the unfolding procedure applied [14, 17, 18]. It was specifically shown that procedures like local unfolding, Gaussian broadening [16], and un-careful polynomial unfolding [17] lead to a false increase of the long-term statistics. The effect of the polynomial order, in polynomial unfolding, on the long-term statistics was studied in [9, 17, 18].

Here, we consider the polynomial unfolding of spectra of ordered eigenvalues of large random matrices. While [9] suggested a detrending process that includes removing some percentage of the eigenvalues prior to the unfolding decreases the effect, [17] suggested the existence of optimal polynomial order for each system.

In this letter we review and reproduce the findings previously reported in [17, 18]. We discuss two special cases for which the spectral density is known. One case is a simple GOE, in which, the levels are obtained from the eigenvalues of the adjacency matrix of a random network. The other has a composite spectrum of independent GOE sequences, obtained from the eigenvalues of a block-diagonal random matrix. The GOE and $m$-GOE systems were chosen because they are well known examples of systems presenting quantum chaos [9], and because the theoretical results for the statistics of chaos are known [1, 12, 19]. Next, we show that the reported dependence on the polynomial order is mainly due to the inclusion of the extreme eigenvalues. Removing them before the polynomial unfolding process results in greatly reducing the effect.

## 2. Background

Random matrix theory (RMT) was proposed by Wigner to explain statistical properties of nuclear spectra [20]. Later, RMT was successfully applied in studying different complex systems, including network spectra. The use of RMT in analyzing complex networks is based on the concept of representing a network as an adjacency matrix. Whole information about a network is encoded in its adjacency matrix. The adjacency matrix is an $N \times N$ matrix, where $N$ represents the number of nodes in a network. For un-weighted undirected networks, where no self connections are allowed, the adjacency matrix is symmetric with 0's on the diagonal. The off-diagonal elements represent links in the network. Their values are either 0 or 1, where 1 represents a link between two nodes, and 0 represents no link present between the two nodes. When the number of 1's in a row follows a Gaussian distribution with mean $p$ and variance

$p(1-p)$, this type of matrix belongs to a GOE, and is very well studied within the RMT framework [21-23]. The mean level density, for a GOE adjacency matrix, follows Wigner's semicircular law;

$$\rho_{GOE}(N,E) = \frac{2N}{\pi a^2}\sqrt{a^2 - E^2}, \text{ for } |E| \leq a \tag{1}$$

where $a$ is the radius of the semi-circle, which is related to the standard deviation $\sigma$ of the off-diagonal elements of the adjacency matrix by

$$a = 2\sigma\sqrt{N}$$

Another case of special interest occurs when studying the interconnections between two or more networks. The overall network formed in such case consists of $m$-clusters of densely connected networks with sparse connections between the clusters. The adjacency matrix in this case is a block-diagonal matrix, i.e. $m$-GOE. Increasing the number of interconnections changes the matrix gradually from the block diagonal form to the form of an adjacency matrix of a single randomly connected network. This model was successfully applied in [24] to study systems in the initial phase of the transition from integrability to chaos.

*2.1 Spectral Statistics*

When studying the spectral properties of systems, two sets of statistics are usually considered. The nearest neighbor spacing distribution (NNSD) is used to study the short-range fluctuations in the spectrum. NNSD is defined as the probability distribution of the spacing between neighboring eigenvalues after unfolding; $p(s)$ where; $s_i \equiv \epsilon_{i+1} - \epsilon_i$. It was shown that the NNSD is insensitive to the unfolding method [17, 18], unlike, the long-term statistics that we discuss next.

The statistical analysis of long-range correlations of level spectra is usually carried out in terms of either the level number variance $\Sigma^2$ or the spectral rigidity $\Delta_3$ [1, 2]. By definition, $\Sigma^2(L)$ measures the variance of the number of eigenvalues in an interval of length $L$ of the unfolded spectrum [25];

$$\Sigma^2(L) = \left\langle (L - n(L,\alpha))^2 \right\rangle_\alpha ; \; L > 0. \tag{2}$$

The Dyson-Mehta $\Delta_3(L)$ statistic (the spectral rigidity) is another quantity which is often used to characterize the long-range correlations in quantum spectra. This is a measure of the average deviation of the spectrum on a given length $L$ from a regular "picket fence" spectrum of a harmonic oscillator [26];

$$\Delta_3(L) = \left\langle \frac{1}{L} \min_{(A,B)} \int_\alpha^{\alpha+L} (N(\varepsilon) - A\varepsilon - B)^2 d\varepsilon \right\rangle \tag{3}$$

The relationship between $\Delta_3(L)$ and $\Sigma^2(L)$ is given [1, 2] as:

$$\Delta_3(L) = (2/L^4) \int_0^L (L^3 - 2L^2 r + r^3) \Sigma^2(r) dr \tag{4}$$

Both statistics are directly related. In the rest of the letter we consider the spectral rigidity.

### 2.2 Extreme eigenvalues

We consider the network as a random graph. In random graph model of Erdös and Rényi, any two nodes are randomly connected with probability $p$ [22]. In the $N \to \infty$ limit, the rescaled spectral density of the uncorrelated random graph converges to the semi-circle law of Eq. (1)

[27]. This limit was shown to be reached asymptotically at $N = 1000$ [17]. After finding the eigenvalues of the adjacency matrix $A$ of the graph, we denote the largest eigenvalue of $A$ by $\lambda_{max}$. Assuming that the graph is connected, $\lambda_{max}$ is unique, real, and positive, by the Perron-Frobenius theorem [28]. Furthermore it was noted that $\lambda_{max}$ is often well separated from the rest of the eigenvalues [29]. It was shown that, as $N$ increases with $p$ fixed, $\lambda_{max}$ scales with $N$, but the circle radius scales with $N^{0.5}$ [30].

For directed graphs, the eigenvalues are distributed as shown in Figure 1(a). In the complex plane, all the eigenvalues are contained in a circle, with the exception of a single (maximum) eigenvalue that is well isolated from the others. When the matrix is symmetric, such is the case with adjacency matrices of un-directed networks, all the eigenvalues collapse to real-valued ones, as shown in Figure 1(b). Figure 1(c) shows the eigenvalues of a 2-GOE matrix, where two eigenvalues are well separated. In the transition from two-clustered network to a single random one, by the introduction and removal of links, one of the two extreme levels increases its separation, while the other one approaches the remainder of the levels, as Figure 1(d) shows.

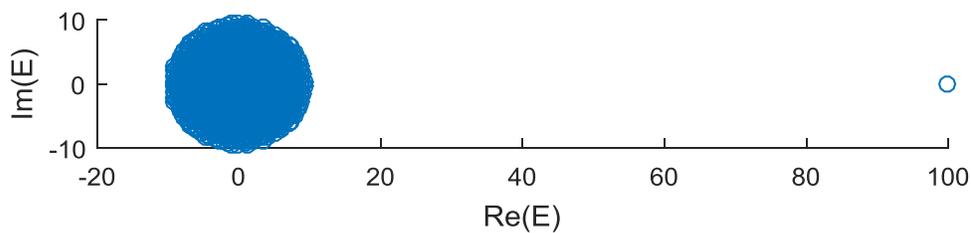

(a)

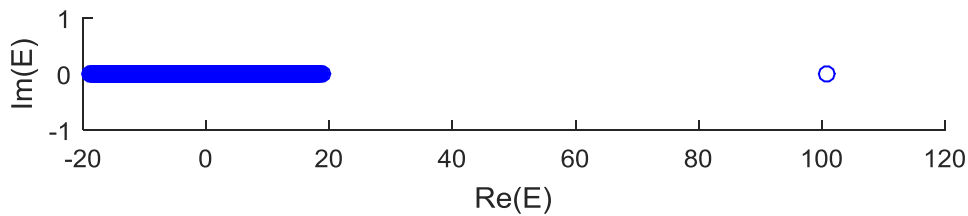

(b)

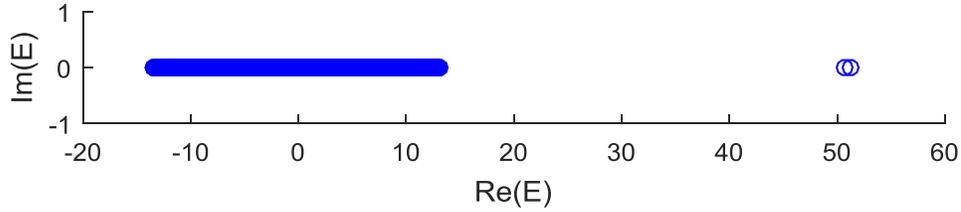

(c)

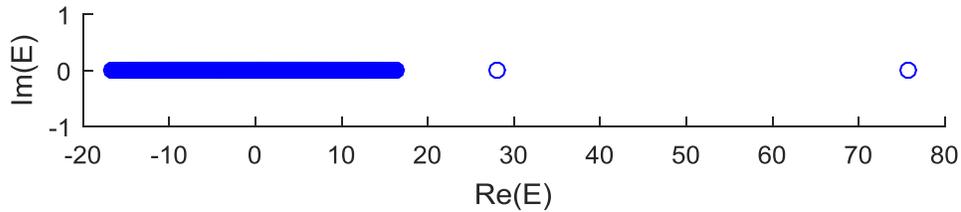

(d)

Figure 1. Complex plane plot of all the eigenvalues of 1000 × 1000 randomly generated adjacency matrices are shown plotted in the complex plane for a case where $N=1000$ and $A_{ij} = 1$ with probability $p = 0.1$ and $A_{ij} = 0$ otherwise. (a) Directed graph, aside from the single largest eigenvalue, all the other eigenvalues are contained within a circle. (b) Adjacency matrix of a single un-weighted un-directed randomly connected network, i.e. GOE. (c) Adjacency matrix of a two clustered un-weighted un-directed randomly connected network, i.e. 2-GOE. (d) Adjacency matrix of a random network that is transiting from two to a single cluster.

The shown distributions of the eigenvalues make it natural to conclude that the largest eigenvalue (the largest '*m*' eigenvalues in m-clustered networks) has a different significance than the others. Intuitively, Their exclusion from the unfolding process would have an impact on the results. This is discussed next.

## 3. Results

In this section we present the $\Delta_3$ statistics for different ensembles of adjacency matrices that model random and clustered networks. For all cases, we constructed ensembles of 20 matrices with $N = 1000$, calculated the spectral characteristics for each matrix separately, and then took the ensemble average for each of the characteristics.

Palla and Vattay [31] have shown that the statistical properties of random networks with mean degree of nodes $k = pN \gg 1$ show GOE statistics. GOE is an example of the systems for which a natural unfolding procedure exists. Because $\rho(E)$ is known to be semicircular for large GOE matrices [32] (i.e. Eq. (1)). In this case;

$$I_{GOE}(N,E) = N\left[\frac{1}{2} + \frac{E}{\pi a^2}\sqrt{a^2 - E^2} + \frac{1}{\pi}\arctan\left(\frac{E}{\sqrt{a^2 - E^2}}\right)\right], \text{ for } |E| \leq a. \quad (5)$$

For a GOE, the spectral rigidity is given by [33];

$$\Delta_3(L) = (Log(L) - 0.0687)/\pi^2$$

And for m-GOE systems;

$$\Delta_{3_{m-GOE}}(L) = m \cdot \Delta_{3_{GOE}}(L/m)$$

Figure 2(a) shows the spectral rigidity $\Delta_3(L)$ for the same ensemble of adjacency matrices for random networks, computed using all eigenvalues, with connectivity probability $p = 0.1$. The different lines show unfolding that was done with different degrees of polynomials compared with the asymptotic formula and the theoretical value for GOE. Figure 2(b) shows the results using the same unfolding procedure, but after excluding $\lambda_{max}$. Figures 3(a) and 3(b) show the results for a 2-GOE system. Figures 2 and 3 show the spectral rigidity for the cases of GOE and

2-GOEs, for the range of $0 < L \leq 200$. As the figures show, when the extreme eigenvalues are removed prior to the polynomial unfolding, the resulting spectral rigidity becomes much less sensitive to the utilized polynomial degrees. The false conclusions about the chaoticity of the system, which can be reached, based on figures 2(a) and 3(a) can no longer be made.

The results show that extra care has to be taken in the unfolding process. when extreme levels are present in the spectrum, they should be removed prior to unfolding. In situation where it is hard to verify the existence of such distinct levels (e.g. by visual examination of the spectrum being studied), the extreme levels may be considered to be outliers, and methods of anomaly detection can be utilized to remove them. For example, in our results, omitting the levels larger than $\langle E \rangle + 3\sigma_E$ (mean level value + three times the standard deviation) resulted in exactly excluding the extreme eigenvalues affecting the results.

It must be noted that, when the largest eigenvalues are included in the unfolding process, increasing the order of the unfolding polynomial does not necessarily guarantee better results. A higher order polynomial would cause the over-fitting phenomena mentioned earlier. This can be seen in figures 2(a) and 3(a), by observing the spectral rigidity obtained after unfolding the spectrum with $25^{th}$ degree polynomial. The values of $\Delta_3(L)$ are less than what they should be for both cases (GOE and 2-GOE). Ref. [17] studied the optimum order of polynomial in such cases and found that it varies with both the matrix size and the number of blocks in block-diagonal matrices.

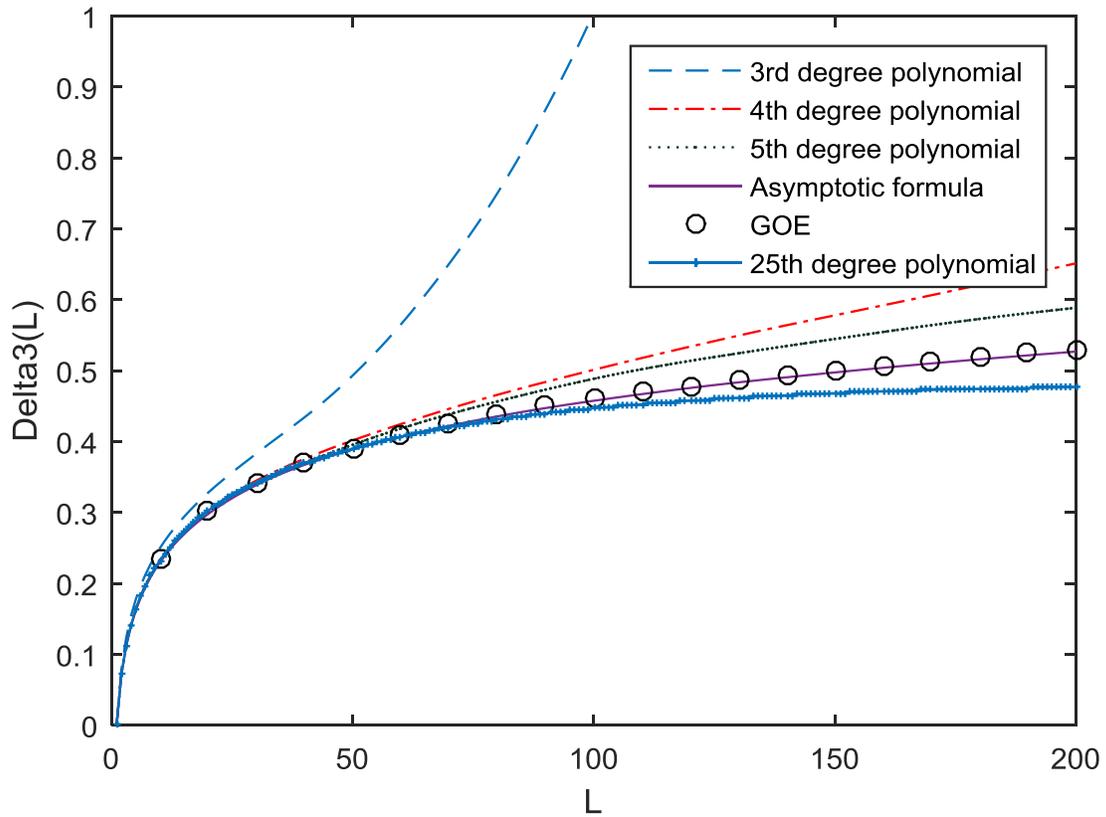

Figure 2(a). Spectral rigidity for GOE (of 20 1000 × 1000 matrices), calculated for 0 < L ≤ 200. The results of unfolding using different degrees of polynomials, and using the asymptotic formula are shown. All eigenvalues are included in the unfolding process.

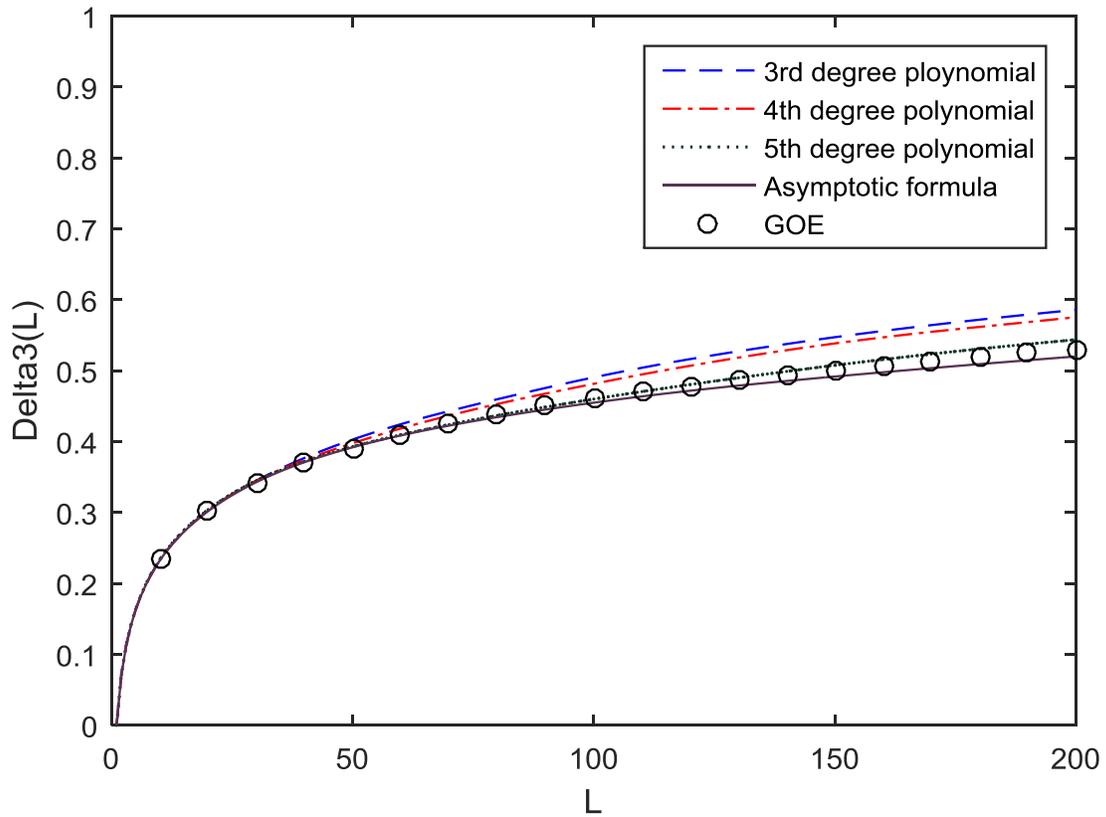

Figure 2(b). Spectral rigidity for GOE (of 20 1000 × 1000 matrices), calculated for $0 < L \leq 200$. The results of unfolding using different degrees of polynomials, and using the asymptotic formula are shown. The largest eigenvalues is excluded from the unfolding process.

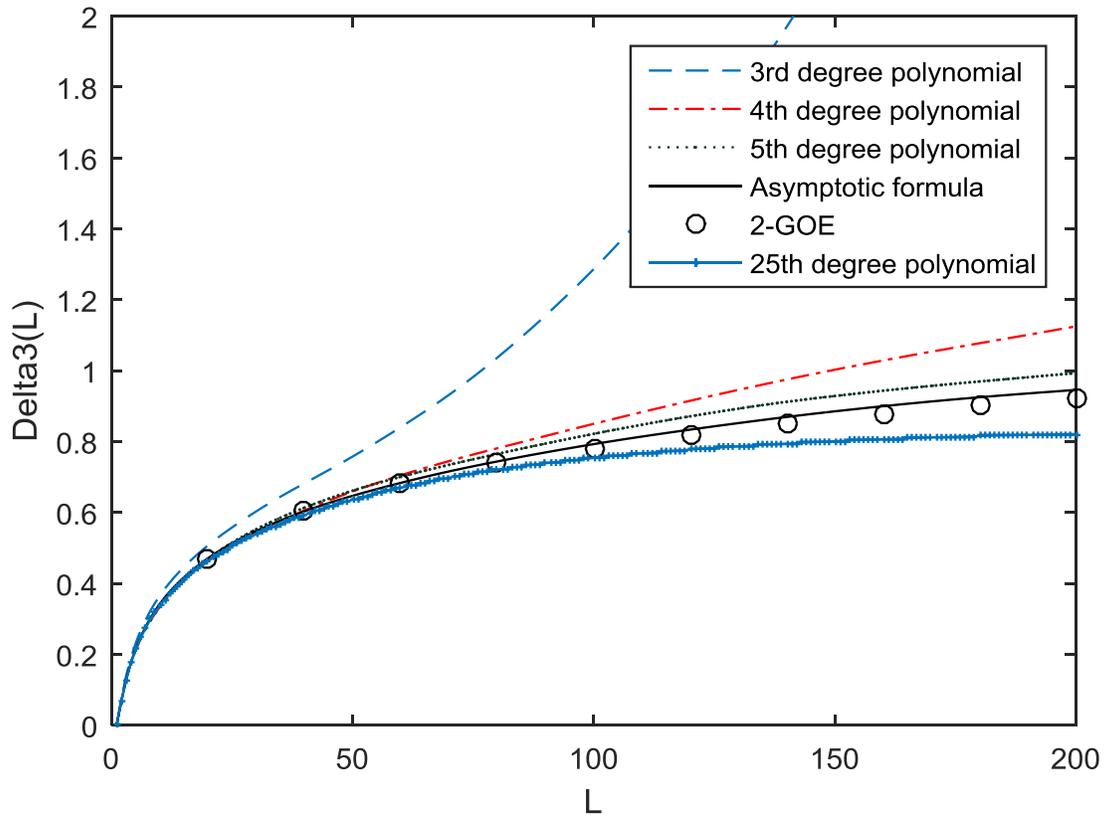

**Figure 3(a).** Spectral rigidity for 2-GOE's (of 20 1000 × 1000 matrices), calculated for $0 < L \leq 200$. The results of unfolding using different degrees of polynomials, and using the asymptotic formula are shown. All eigenvalues are included in the unfolding process.

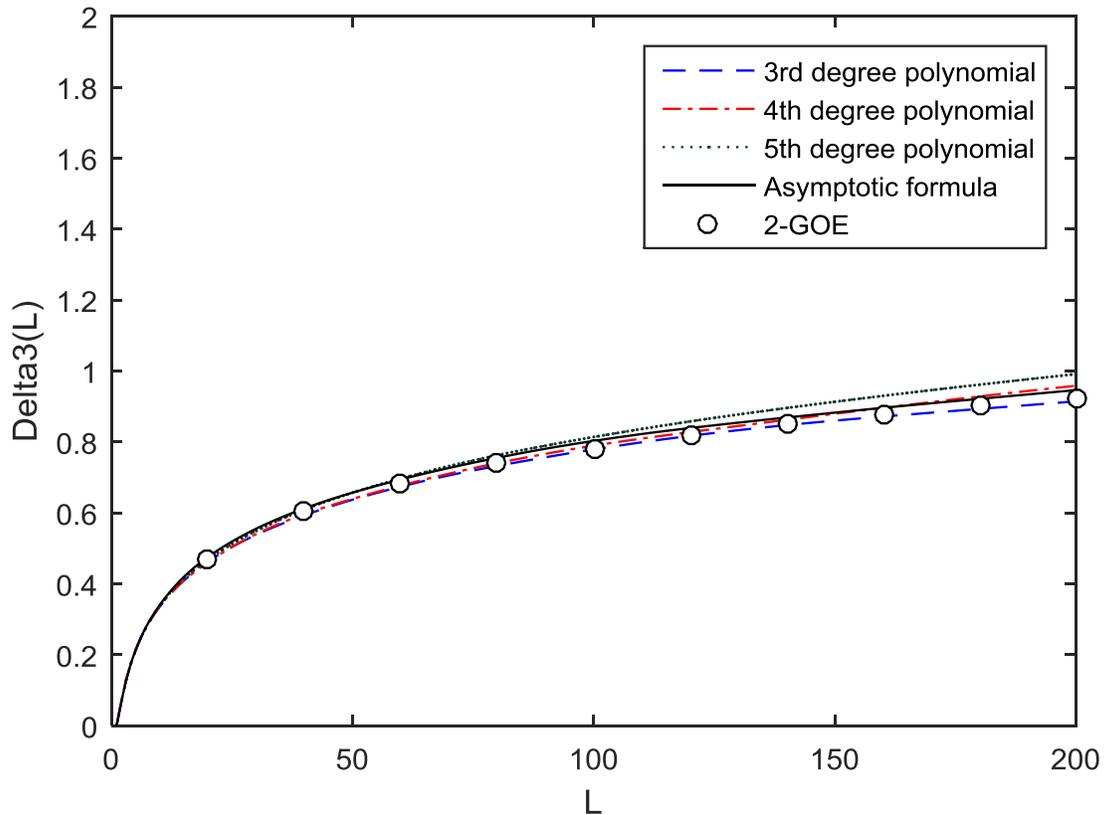

**Figure 3(b). Spectral rigidity for 2-GOE's (of 20 1000 × 1000 matrices, calculated for 0 < L ≤ 200). The results of unfolding using different degrees of polynomials, and using the asymptotic formula are shown. The two largest eigenvalues are excluded from the unfolding process.**

Figure 4 shows why the inclusion of the extreme eigenvalue(s) leads to wrong unfolding. In figure 4(a), for the fitted polynomials to pass through the extreme eigenvalue they deviated away from following the trend of the staircase function. The removal of the extreme value enabled the fitted curves to resemble the staircase function. The insets of the figures show the deviation of the fitted curves from the staircase function at the beginning of the curve.

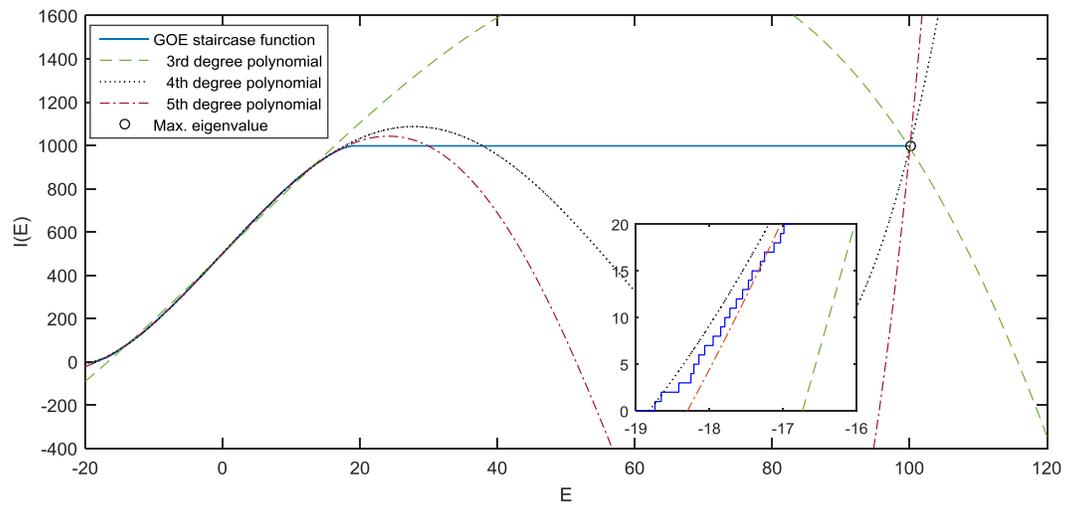

Figures 4(a) Polynomial fitting to the GOE staircase function, with the inclusion of the extreme eigenvalue. The inset shows the deviation of the fitted curve from the staircase function at the beginning of the curve.

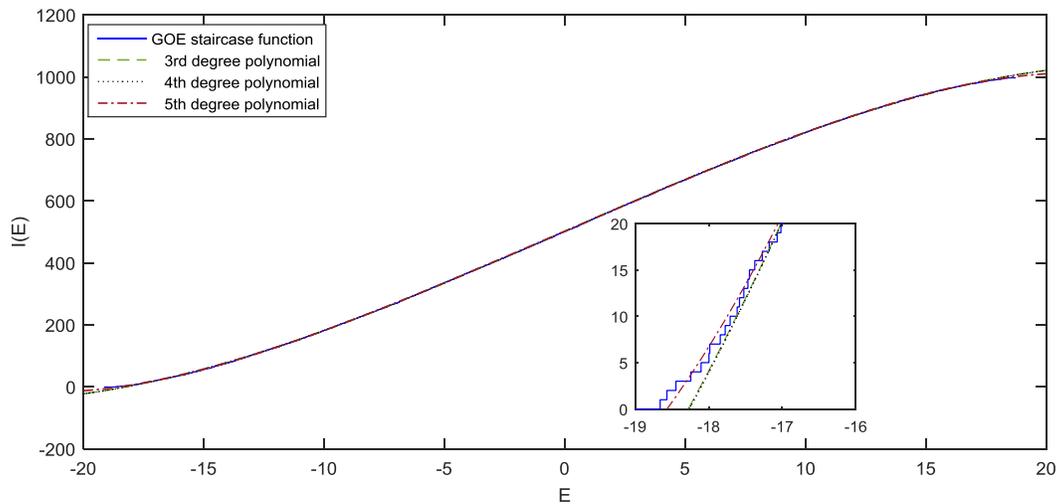

**Figures 4(b) Polynomial fitting to the GOE staircase function, after removing the extreme eigenvalue. The inset shows the deviation of the fitted curve from the staircase function at the beginning of the curve.**

## 4. Conclusions

This letter revisited the reported problem of the sensitivity of the long-range statistics on the polynomial degree in polynomial unfolding of the spectrum of adjacency matrices of complex networks. The two considered cases of a randomly connected network and a clustered network are well known examples that represent chaotic and mixed systems. We showed that, contrary to previous conclusions, it is not the degree of the polynomial that has the main impact on the estimated long-term correlation behavior, but rather the inclusion of the extreme eigenvalues in the polynomial unfolding. For $m$-block diagonal adjacency matrices, there exist $m$ extreme eigenvalues that are well separated from the remaining eigenvalues. Removal of such eigenvalues, prior to the unfolding process, greatly reduces the previously reported impact. The false conclusions about the chaoticity of the system can no longer be made regardless of the unfolding polynomial degree. Results show that this is even more apparent in the case of composite spectra (i.e. 2-GOE).